\documentclass[twocolumn]{aastex631}
\usepackage{ulem} 
\usepackage{cancel}

\newcommand{\gr}{$\gamma$-ray}
\newcommand{\sg}{SG0837}
\newcommand{\fermi}{{\it Fermi}}

\shortauthors{Zhang and Xin}
\graphicspath{{./}{figures/}}

\begin{document}

\title{Discovery of an Extended $\gamma$-ray Emission around the Supernova Remnant Candidate associated with PSR J0837$-$2454}

\author{Pengfei Zhang}
\affiliation{Department of Astronomy, School of Physics and Astronomy, Key Laboratory of Astroparticle Physics of Yunnan Province, Yunnan University, Kunming 650091, People's Republic of China; zhangpengfei@ynu.edu.cn}

\author{Yuliang Xin}
\affiliation{School of Physical Science and Technology, Southwest Jiaotong University, Chengdu 610031, People's Republic of China; ylxin@swjtu.edu.cn}

\begin{abstract}
Motivated by the recent discovery of a low surface brightness diffuse emission,
a supernova remnant (SNR) candidate, surrounding the young pulsar PSR~J0837--2454,
we carry out a likelihood analysis of the $\gamma$-ray data obtained by the \emph{Fermi} Gamma-ray Space Telescope
from August 2008 to November 2022. Using a 2D Gaussian spatial template,
we detect a significant extended $\gamma$-ray emission with a 68\% containment radius of $\sim1^{\circ}.8$, 
which is spatially coincident with the new SNR candidate at $\sim12\sigma$ confidence level.
The spectrum of the extended $\gamma$-ray emission, obtained in the energy range of 0.1–500.0 GeV,
shows a significant spectral curvature at $\sim$1 GeV, with a log-parabola spectral shape.
Several scenarios, such as the SNR, pulsar wind nebula, and pulsar halo,
are discussed as the potential origins of the extended $\gamma$-ray emission,
and our model fitting results are preferred for the SNR scenario.

\end{abstract}

\keywords{Gamma-rays(637); Pulsars (1306); Supernova remnants (1667)}

\section{Introduction}
\label{Intro}
During the ending stage of massive star evolution, the core of the star may undergo a powerful
supernova explosion, collapsing into a rotating neutron star (i.e., a pulsar),
which may lead to the creation of a supernova remnant (SNR),
as the expanding gaseous remnant interacts with the surrounding circumstellar and interstellar medium.
In our Galaxy, nearly 300 SNRs have been identified by the radio observations \citep{g14,g19}
at low Galactic latitudes $\lesssim$ 300 pc \citep{m01}.
Thanks to the $\gamma$-ray telescopes, approximately 40 SNRs have been detected with $\gamma$-ray
emissions \citep[][and references therein]{zxl19}, including the GeV $\gamma$-ray SNRs detected by {\em Fermi},
e.g. IC 443 and W44 \citep{2013Sci...339..807A},
and the TeV $\gamma$-ray SNRs detected by the ground-based Cherenkov telescopes (e.g. HESS, HAWC, VERITAS and LHAASO), 
such as RX J1713.7-3946 \citep{2018A&A...612A...6H}, G106.3+02.7 \citep{2009ApJ...703L...6A,2020ApJ...896L..29A,2021Natur.594...33C}, etc.
SNR's electromagnetic emissions extend from MHz radio frequencies to TeV $\gamma$-ray energies \citep{zxl19,zxz+21}.
The high-velocity shock of SNR could accelerate cosmic rays to very high energies
(even up to hundred of TeV).
The studies of the $\gamma$-ray emissions from SNRs provide us excellent tools for probing
the interstellar medium and stellar evolution in our Galaxy, especially for the Galactic cosmic
rays acceleration.

Recently, \citet{pbh+21} reported a discovery and timing of a young pulsar PSR J0837--2454
at a high Galactic latitude with a Galactic coordinate (J2000) of $l=247^{\circ}.6$ and $b=9^{\circ}.8$.
They presented the pulsar's timing solution by using the radio data from the Parkes radio telescope.
Its spin period ($P$) and spin-down rate ($\dot{P}$) are 629.4 ms and 3.5$\times10^{-13}$ s s$^{-1}$, respectively.
And the characteristic age \citep[$\tau_{\mathrm{c}}=\frac{P}{2 \dot{P}}$;][]{hpa12} is 28.6~kyr
based on the assumption that the magnetic-dipole braking as the
only energy-loss mechanism with a braking index of 3 and $P\ll P_{\rm init}$ ($P_{\rm init}$ is
the pulsar's initial spin period).
Its spin-down luminosity ($\dot{E}$) is calculated to be $5.5\times10^{34}$ erg s$^{-1}$,
and the surface dipole magnetic field strength ($B_{\rm S}$) is $1.5\times10^{13}$~G.
Based on the NE2001 electron density model provided in \citet{cl02},
\citet{pbh+21} claimed that the pulsar locates at a larger distance of 6.3 kpc inferred by a DM-derived distances,
which implies that PSR J0837--2454 appears at the edge of Galaxy and has a z-height
above the Galactic plane of 1.1 kpc.
If this value is true, PSR J0837--2454 will be the first pulsar known to be born from a runaway O/B star.

Furthermore, they also claimed a discovery of a low surface brightness diffuse emission with
a region of $\sim1^{\circ}.5$ extent concentrated around PSR~J0837--2454 by using the archival Galactic
and Extragalactic All-sky Murchison Widefield Array Survey \citep[GLEAM;][]{GLEAM+17} data
in 170--231 MHz bands\footnote{https://vo.astron.nl/tgssadr/q\_fits/cutout/form}.
And the diffuse emission has a morphology consistent with a SNR.
Based on the data from GLEAM and Southern H$\alpha$ Sky Survey Atlas
\citep[SHASSA;][]{shassa01}, the distance for the diffuse emission is estimated to be $\sim$0.9
and 0.2~kpc, respectively, which is much smaller than that predicted by \citet{cl02}.
If the diffuse emission is indeed an SNR associated with the high Galactic latitude pulsar, searching for the multi-wavelength emission of the SNR, especially in $\gamma$-ray, would be helpful to study the particle acceleration and probe the interstellar medium above the Galactic plane.
Motivated by their report, we carried out the data analysis with the $\gamma$-ray data surrounding
PSR~J0837--2454 collected by the Large Area Telescope onboard the \emph{Fermi} Gamma-ray Space Telescope \citep[\emph{Fermi}-LAT;][]{lat09}.
And this paper is structured as follows: the likelihood analysis for the \emph{Fermi}-LAT
data and the main results are described in Section~\ref{sec:lat-data}. Discussions of the probable
physical origins for the extended $\gamma$-ray emission are shown in Section~\ref{sec:dis}.
In Section~\ref{sec:sum}, we present a summary.


\section{Data Analysis and Results}
\label{sec:lat-data}
\subsection{\fermi-LAT Data and source model}
\label{sec:model}
Around the position of PSR~J0837--2454 reported by \citet{pbf+17} (R.~A.=$129^{\circ}.49$, decl.=$-24^{\circ}.91$),
there is a $\gamma$-ray point source named as 4FGL~J0838.9--2502 in the Data Release 3 of
the fourth Fermi-LAT source catalog \citep[4FGL-DR3;][]{dr3+22} based on 12 yr data.
In the 4FGL-DR3, 4FGL J0838.9-2502 has no associated source in other wavelengths,
and its $\gamma$-ray spectrum is described by a log-parabola spectral shape (LP)
of $dN/dE=N_0(E/E_b)^{-[\alpha+\beta \log (E/E_b)]}$ with $\alpha$=2.60, $\beta$=0.45 and $E_{b}$=1.03 GeV \citep{dr3+22}.
The following data analysis is aiming to determine the association between this $\gamma$-ray source and
PSR J0837-2454 or the SNR around it.



Firstly, we carried out a whole data analysis in order to update the catalog's parameters for
the $\gamma$-ray sources in the region of interest (RoI) with the 14 yr \emph{Fermi}-LAT observations.
We selected the \emph{Fermi}-LAT Pass 8 \textit{Front+Back} events (evclass = 128 and evtype = 3)
in the energy range of 0.1--500.0 GeV within a $20^\circ\times20^\circ$ RoI
centered at the position of 4FGL~J0838.9$-$2502.
The observations span from August 4 2008 to November 24 2022 (MJD: 54682.687-59907.214).
The events with zenith angles $\geqslant90^{\circ}$ were removed to exclude the $\gamma$-ray contamination
from the Earth Limb. The expression of ``DATA\_QUAL $>$ 0 \&\& LAT\_CONFIG == 1" was used,
for \emph{gtmktime}, to save the events having high-quality with flags of ``good" in the good time intervals.
In our data analysis, the instrumental response function of “P8R3\_SOURCE\_V3”
and the software package of Fermitools–2.2.0 were used for the data reduction.

\begin{figure*}
\centering
\includegraphics[angle=0,scale=0.65]{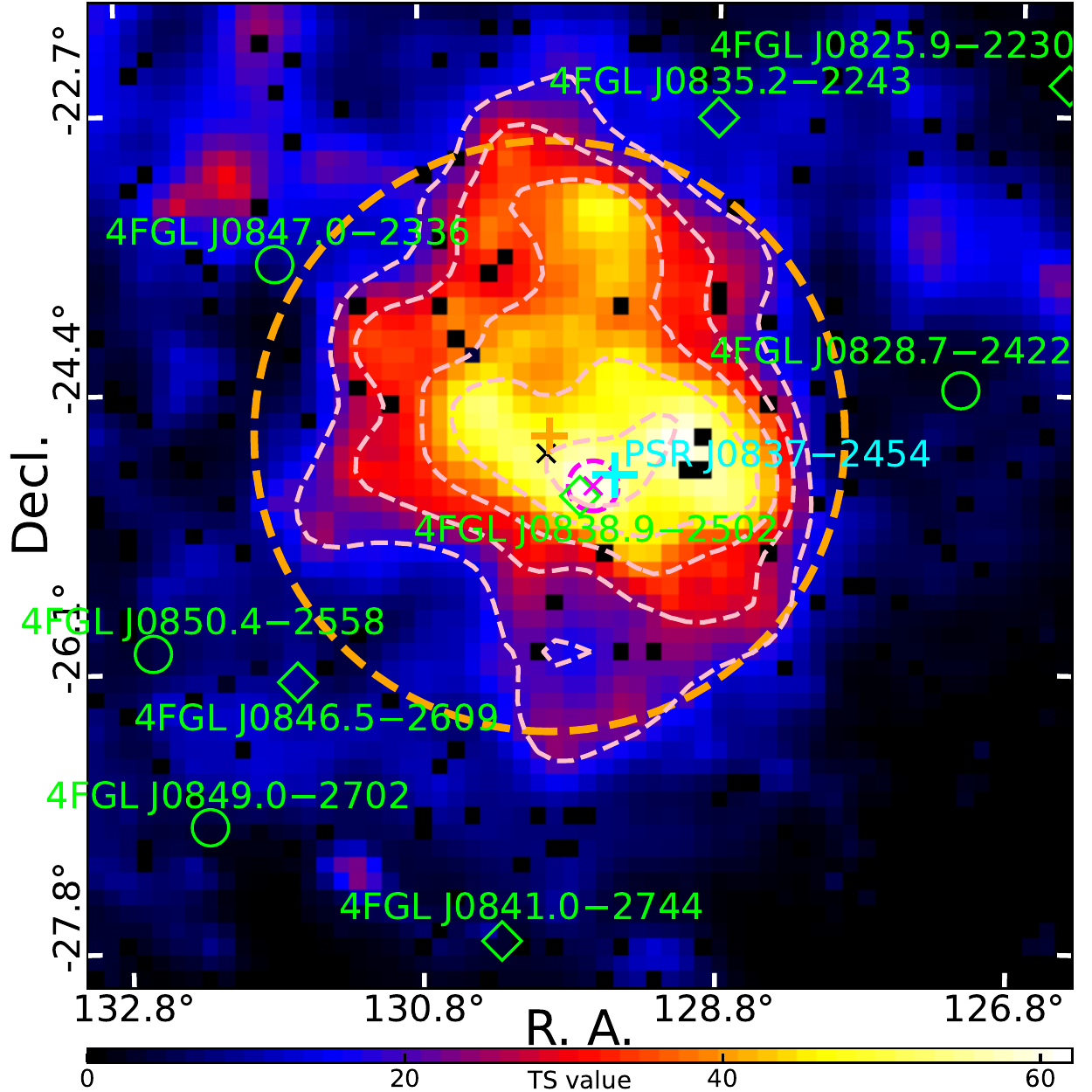}
\includegraphics[angle=0,scale=0.65]{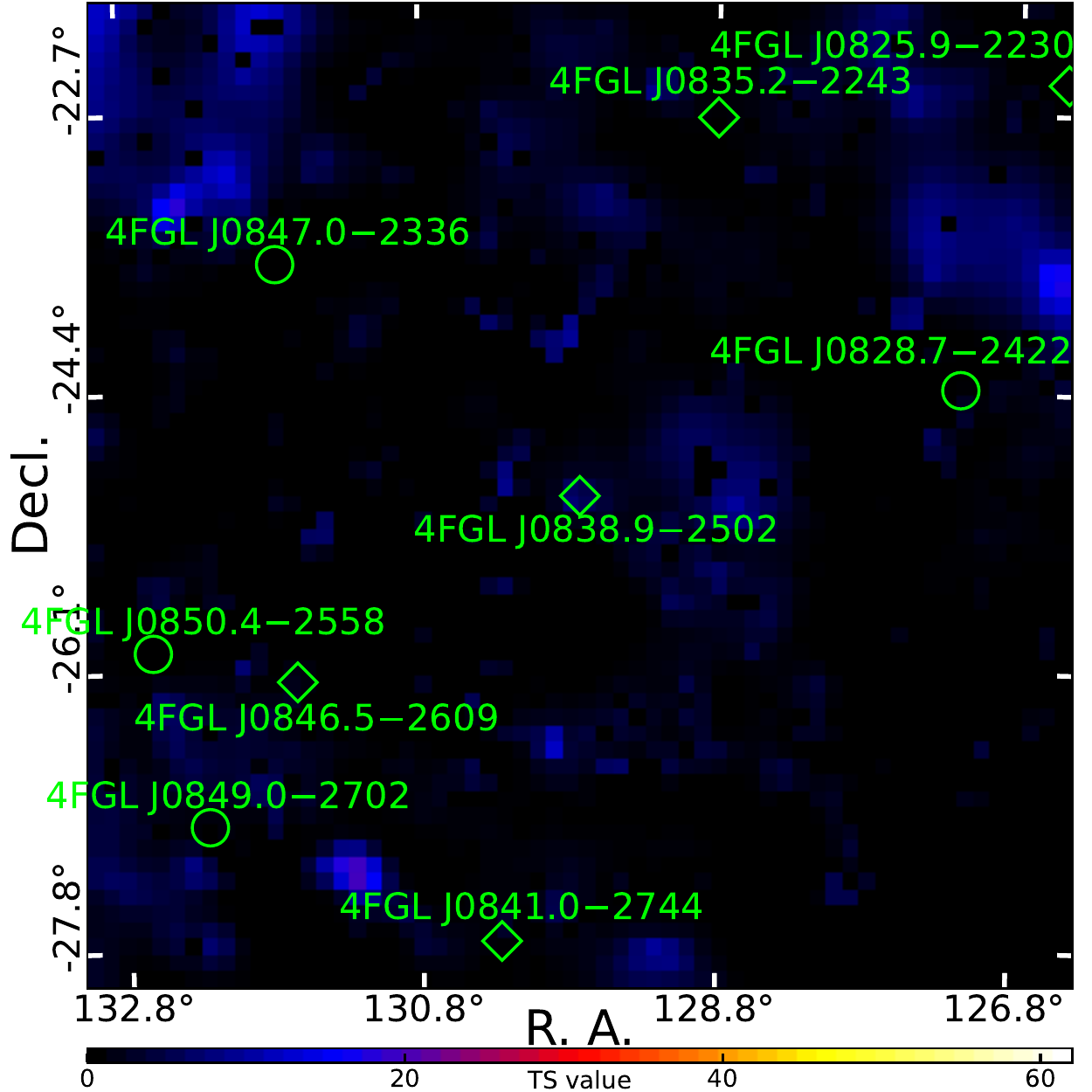}
\caption{TS maps in 0.5--500.0 GeV covering $6^{\circ}\times6^{\circ}$ region around the new found \sg,
        centered at the position of 4FGL J0838.9--2502 with each pixel of $0^{\circ}.1$.
        Left panel: TS map for the $\gamma$-rays from \sg. PSR J0837--24 is flagged with a cyan plus.
        The best-fit \emph{Fermi}-LAT position of the 2D Gaussian template
        is marked with an orange plus, and its best-fit extension is indicated by an
        orange dashed circle. The best-fit positions of the uniform disk and point source
        models are marked with black and purple crosses, respectively. The positional uncertainty
        of point source is marked by the purple dashed circle. The pink contours are derived based on
        the TS value at each pixel, and the $\Delta$TS between the neighbouring contours is 8.6.
        All \gr~sources in the 4FGL are colored with green and labeled with `4FGL'. Right panel: the residual TS map by subtracting the $\gamma$-rays from all sources inlcuding \sg.
        And the two TS maps share same scaled colorbar for convenient comparison.}
\label{fig:map}
\end{figure*}

Based on the newest 4FGL-DR3 catalog, we used a python script,
make4FGLxml.py\footnote{http://fermi.gsfc.nasa.gov/ssc/data/analysis/user/}, to create a model file.
The model file includes all the spectral parameters of the $\gamma$-ray sources in 4FGL-DR3
within 25$^\circ$ around 4FGL~J0838.9$-$2502. We freed the normalizations and spectral parameters for
the sources within 5$^{\circ}$ of the ROI center, and the normalizations for the sources
within 5$^{\circ}$--10$^{\circ}$, together with the ones which are 10$^{\circ}$ outside but identified as variable sources.
The normalizations of Galactic and extragalactic diffuse emission components were also set free.
All other parameters in the model file were fixed to be their values provided in 4FGL-DR3.
Then, a binned maximum likelihood analysis was
performed between the whole LAT data set and the above model file. 
Then we saved all the best-fit parameters as a new model file.
In order to reveal the $\gamma$-ray emissions around 4FGL~J0838.9--2502,
a TS map with a region of $6^{\circ}\times6^{\circ}$ was created based on the new model with $gttsmap$
by fixing all model parameters for all 4FGL-DR3 sources including the
two diffuse backgrounds, and removing 4FGL~J0838.9--250 from the model.
The TS map is shown in the left panel of Figure~\ref{fig:map}, and a significant extended $\gamma$-ray emission
(hereafter named as \sg) is positional coincident with the SNR candidate reported in \citet{pbh+21}.

\subsection{Spatial Analysis}

\begin{table*}
\begin{center}
\caption{Spatial Analysis Results}
\scriptsize
\begin{tabular}{ccccccc}
\hline\hline
Template$^{(1)}$ & Position$^{(2)}$ & TS$^{(3)}$ & $\Gamma^{(4)}$ & --$\mathcal{L}^{(5)}$ & DoF$^{(6)}$ \\
\hline
Point source & $129^{\circ}.64\pm0^{\circ}.06$, --$24^{\circ}.97\pm0^{\circ}.06$ & 82.44 & $2.54\pm0.09$ & --1018839.61 & 4\\
Uniform disk & $129^{\circ}.95\pm0^{\circ}.09$, --$24^{\circ}.78\pm0^{\circ}.14$; $R_{68}=1^{\circ}.97^{+0^{\circ}.25}_{-0^{\circ}.23}$ & 161.30 & $2.09\pm0.05$ & --1018878.54 & 5\\
Uniform disk+Point source & --- &  158.98; 11.09 & $2.07\pm0.09$; $2.33\pm0.40$ & --1018882.81 & 9 \\
2D Gaussian  & $129^{\circ}.93\pm0^{\circ}.14$, --$24^{\circ}.67\pm0^{\circ}.14$; $R_{68}=1^{\circ}.80^{+0^{\circ}.25}_{-0^{\circ}.22}$ & 178.88 & $2.06\pm0.04$ & --1018886.21 & 5\\
2D Gaussian+Point source & --- &  176.81; 9.81 & $2.04\pm0.06$; $2.34\pm0.29$ & --1018890.01 & 9\\
\hline\hline
\end{tabular}
\label{tab:ext}
\end{center}
	{\textbf{Notes.} Likelihood results of spatial analysis. (1) Spatial templates used in the spatial analysis.
         (2) Best-fit position and the radius containing 68\% of the intensity ($R_{68}$) derived with
         package of \emph{Fermipy}. The best-fit likelihood results of TS value, PL index,
         and --log(Likelihood) values are listed in Columns (3), (4), and (5), respectively.
         (6) Degree of freedom (DoF) for source models.}
\end{table*}

In order to study the spatial extension of the $\gamma$-ray emission from \sg, we employed the \emph{Fermi}
ScienceTools packaged in Python (\emph{Fermipy}) to derive the best-fit position based on
the assumption that the \sg~is a point source. Its coordinate was derived to be
R.~A.=$129^{\circ}.64$ and decl.=$-24^{\circ}.97$ with a $2\sigma$ error radius of $0^{\circ}.15$,
which is shown as a purple cross in Figure~\ref{fig:map}, together with its error circle marked by the dashed circle.
Based on the new coordinate, we obtained \sg's TS value to be $\sim$82.44 as a point source.
Then we used two spatial models, an uniform disk and a 2D Gaussian extended template,
to describe the $\gamma$-ray emission from \sg. The best-fit central positions
and extensions of the two extended templates are listed in Table~\ref{tab:ext}. 

Using the positions and extensions of the two spatial models, we performed the likelihood analysis again, 
and the best-fit results are summarized in Table~\ref{tab:ext}.
In these analysis, the spectral shape of power-law (PL), $dN/dE=N_0(E/E_0)^{-\Gamma}$,
for each extended spatial model is employed in order to compare their likelihood results easily.
The significance of the extension for a $\gamma$-ray source is defined by a likelihood-ratio test
as shown in \citet{laa+12}, which can be calculated by
TS$\rm_{ext}$ = $-2(\log\mathcal{L}_{\rm pt} - \log\mathcal{L}_{\rm ext})$,
where $\mathcal{L}_{\rm pt}$ (null hypothesis) and $\mathcal{L}_{\rm ext}$ (alternative hypothesis)
are the maximum likelihood values for point source model and spatial extended source model, respectively.
For the likelihood results listed in Table~\ref{tab:ext}, 
we found that the 2D Gaussian template is significantly preferred to the point source model with TS$\rm_{ext}\sim$ 93, which corresponds to the significance level of $9.6\sigma$ with one additional degree of freedom (DoF).

The TS value of the 2D Gaussian template is 178.88 in the likelihood analysis based on the spectral shape of PL,
corresponding to a significance level at $\sim12\sigma$ with five DoFs.
The best-fit position of the 2D Gaussian template was derived to be R.~A.=129$^{\circ}$.93
and decl.=$-24^{\circ}$.67 with a 68\% containment radius of $\sim1^{\circ}.8$.
The position and extension are shown in Figure~\ref{fig:map} with an orange plus and a dashed circle,
respectively. From Figure~\ref{fig:map}, we can see that the angular radius visually encloses
the most of extended $\gamma$-ray emissions from \sg.
We updated the model file with the best-fit values in this likelihood analysis with the 2D Gaussian template.
Then a residual TS map was created based on the updated model by fixing all model parameters for
all sources (including \sg) in the model file, which is shown in the right panel of Figure~\ref{fig:map}.
None of obvious excess suggests that 2D Gaussian template can well describe the \sg's $\gamma$-ray emission.

Meanwhile, we also tested other complex models, i.e. 2D Gaussian/uniform disk plus a point source,
and the likelihood results are shown in Table \ref{tab:ext}.
The TS of the point source is not significant with TS$_{_{\rm PS}} < 16$,
and these models are not favored in the further analysis.
Moreover, we also performed a timing analysis by folded the \emph{Fermi}-LAT data using the ephemeris
for PSR J0837--2454 \citep{pbh+21}, while no creditable pulsation was found.
These analysis make PSR J0837--2454 to be a radio loud and $\gamma$-ray quiet pulsar.

\subsection{Spectral Analysis}
\label{sec:spec}
To test \sg's spectral properties in $\gamma$-rays, \textbf{we used} a spectral form
of LP to fit the $\gamma$-rays from \sg~in 0.1--500.0~GeV. 
And the corresponding TS$_{\rm LP}$ value of \sg~is calculated to be $\sim$198.
Other parameters $\alpha$, $\beta$, and $E_{\rm b}$ are fitted to be $2.23\pm0.22$, $0.29\pm0.09$, and $1.4\pm0.4$~GeV, respectively. 
The corresponding integrated photon and energy fluxes are $(1.36\pm0.21)\times10^{-8}$~photons~cm$^{-2}$~s$^{-1}$
and $(1.26\pm0.18)\times10^{-11}$~erg~cm$^{-2}$~s$^{-1}$, respectively.
The variation of TS value between LP and PL models is $\Delta\rm TS=TS_{LP} - TS_{PL}\sim19$, corresponding to a significance level of $\sim 4.4\sigma$.
Hence we suggest that the LP model is relatively better to the PL one to describe the gamma-ray emission from \sg~.

\begin{figure}
\centering
\includegraphics[angle=0,scale=0.56]{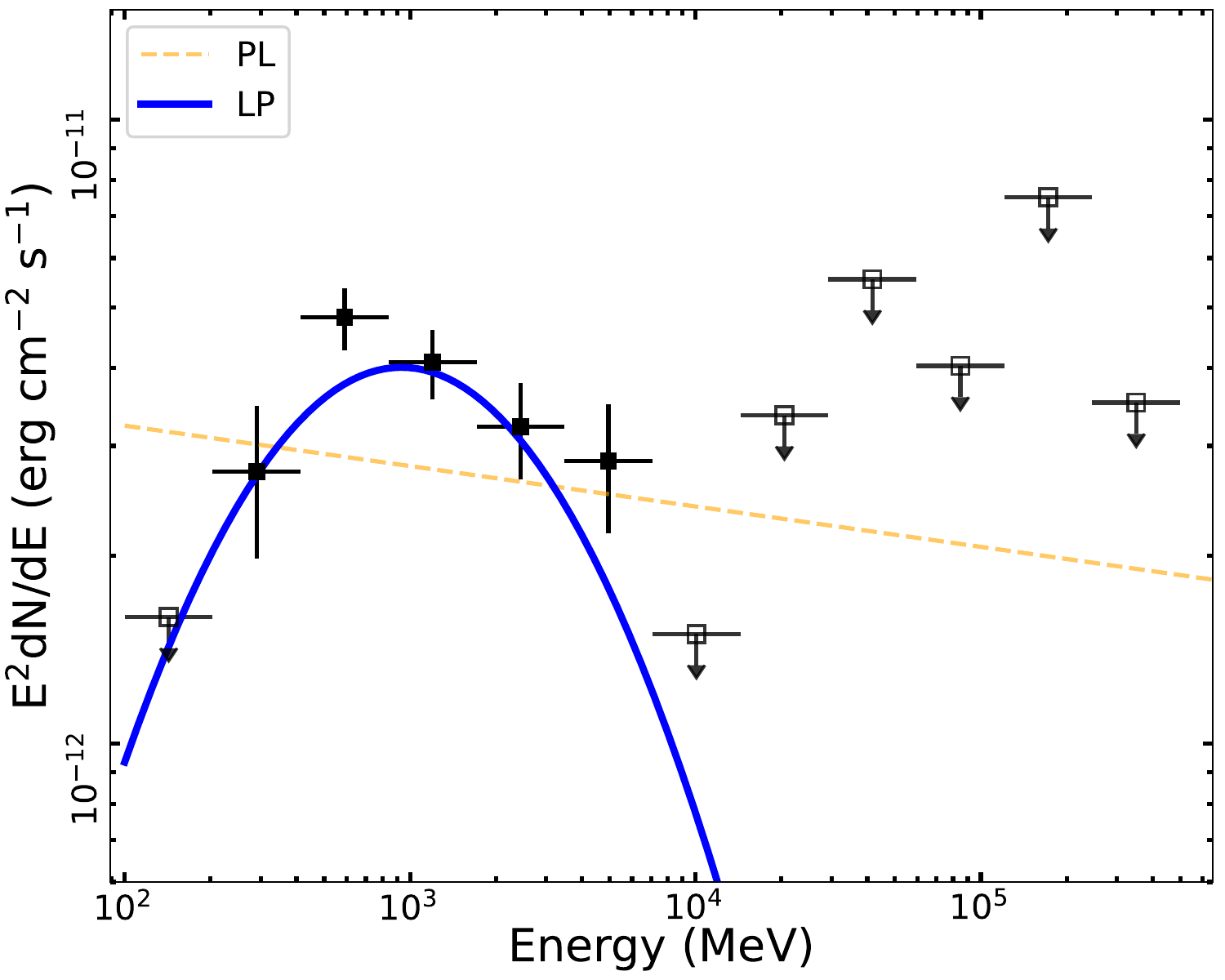}
\caption{\gr~SED of \sg~obtained from the data in 0.1--500 GeV. The best-fit LP and PL spectral shapes are shown
              as the blue solid and yellow dashed lines, respectively. The flux data points with TS $>10$ are shown
              with the black squares with the pluses as their uncertainties, and the black arrows indicate the 95\%
              upper limits.}
\label{fig:spec}
\end{figure}

Then we saved the best-fit parameters into a final model file,
and fixed the spectral parameters for all the sources in the model file
to be the above likelihood analysis values. The normalizations for the sources
within 10$^{\circ}$ around the ROI center and the two diffuse backgrounds were
left free. Based on the final model file, we extracted a spectral energy distribution
(SED) for \sg~in the energy range of 0.1--500.0~GeV by adopting the spatial template of
the 2D Gaussian and the global spectral shape of LP.
The data were divided into 12 equal logarithmically spaced energy bins,
and the individual likelihood analysis was employed for each bin.
We show the $\gamma$-ray spectrum of \sg~ in Figure~\ref{fig:spec},
in which the 95\% flux upper limit is calculated for the energy bin with TS value of \sg~lower than 10.
The global fitting with LP and PL models are also plotted with the blue solid and
yellow dashed lines, respectively.
Comparing the best-fit LP and PL models in Figure~\ref{fig:spec}, \sg's $\gamma$-ray
SED in 0.1-500.0 GeV is relatively well described by the LP model,
which is in agreement with the result shown by $\rm\Delta{TS}$.

\section{Possible Gamma-ray Origins}
\label{sec:dis}

\subsection{Supernova Remnant}

\begin{figure}
\centering
\includegraphics[angle=0,scale=0.68]{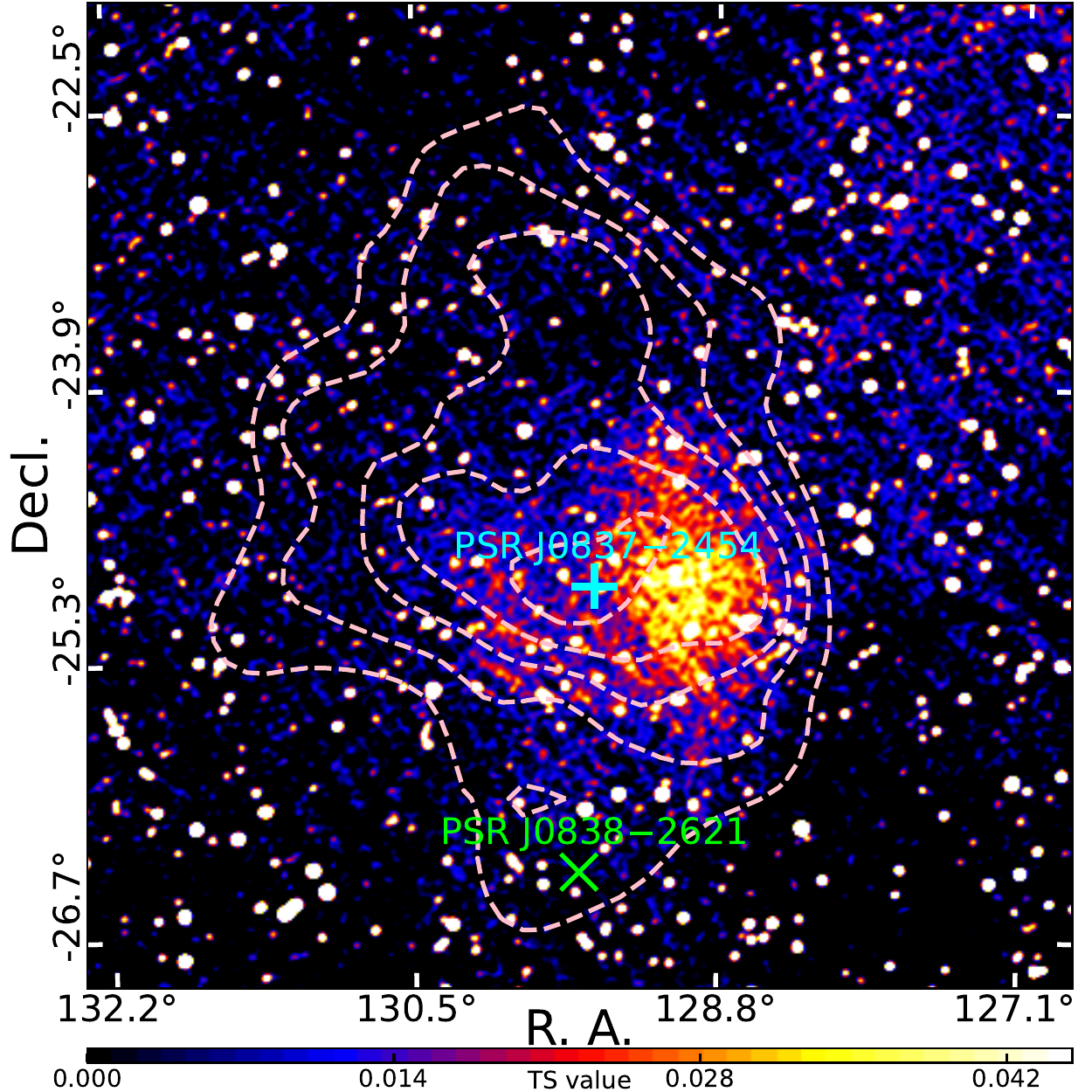}
\caption{Radio imaging with a wide-field view around \sg~in the stacked 170--231 MHz band data.
         The archival data are obtained from the TIFR GMRT sky survey,
         as reported in Figure~6 of \citet{pbh+21}.
         PSR J0837--2454 is flagged with a plus, while the cross stands for the position of
         PSR J0838--2621 with a characteristic age of $1.3\times10^5$~kyr \citep{bjd+06}.
         The pink contours show the GeV $\gamma$-ray emission of \sg~ drawn from Figure~\ref{fig:map}.}
\label{fig:radio}
\end{figure}

According to the discovery in \citet{pbh+21}, thus the origin of the SNR scenario for the $\gamma$-ray emission
from \sg~is considered.
The $\gamma$-ray spectra of tens of {\em Fermi}-LAT observed SNRs can be basically divided into two classes \citep{zxl19}: one has the hard GeV $\gamma$-ray spectrum with the spectral curvature at $\sim$TeV, which corresponds to the young-aged SNR, like RX J1713.7-3946 \citep{2018A&A...612A...6H} or RX J0852.0-4622 \citep{2018A&A...612A...7H}.
And the $\gamma$-ray emissions from these SNRs are typically suggested to be from inverse Compton scattering (ICS) of accelerated electrons (leptonic process). 
Another class shows the spectral break at $\sim$GeV, which corresponds to the old-aged SNRs interacting with molecular clouds, 
like IC 443 and W44 \citep{2013Sci...339..807A}. And such $\gamma$-ray emissions are suggested to be from the decay of neutral pions produced by the inelastic proton-proton collisions (hadronic process). 
The $\gamma$-ray spectrum of SG0837 is similar to that of IC 443 and W44, etc, and the hadronic model is also considered here for it.
Considering the observational fact that the size of the $\gamma$-ray emission region
is much larger than that of the remnant, which is shown as in Figure~\ref{fig:radio}, 
the escaping scenario of protons is suggested, 
i.e. the $\gamma$-ray emission is produced by the protons accelerated and escaped from 
the shock of SNR,
like the SNR W28 \citep{2008A&A...481..401A,2018ApJ...860...69C}.
Here we assume instantaneous injection of protons into an uniform emission zone at $T$ = 28.6~kyr.
Here the age of remnant is assumed to be the characteristic age of PSR J0837-2454 \citep{pbh+21}.
The spectrum of injected protons is adopted to be a power-law with an exponential cutoff $E_{\rm cut}$: 
\begin{equation}
Q_{\rm inj}(E) = Q_{0} E^{-\Gamma} exp(-E/E_{\rm cut}).    
\label{eq:inpr}
\end{equation}
Here the spectral index and cutoff energy of protons are adopted to be $\Gamma = 2.0$ and $E_{\rm cut}$ = 3~PeV, respectively.
And the total energy of injected protons is assumed to be $W_{\rm p, inj}$ = $\eta E_{\rm SN}$,
where $\eta$ is the fraction of the kinetic energy of SNR, $E_{\rm SN}$, converted into the escaped proton energy, 
and the typical value of $E_{\rm SN}$ is adopted to be 10$^{51}$~erg \citep{2005NatPh...1..147W,2006MNRAS.370L..14V}.

The proton spectrum within the emission region can be derived as \citet{2020ApJ...897L..34L}:
\begin{small}
\begin{equation}
N_p(E, t) = \frac{Q(E)}{[4 \pi D(E) T]^{\frac{3}{2}}} \int_{0}^{R} 4\pi r^2 dr \ \rm exp\left[-\frac{r^2}{4D(E)T}\right].
\label{eq:prsp}
\end{equation}
\end{small}
And the diffusion coefficient of protons is assumed to be spatially consistent and energy dependent with $D(E)=\chi D_0(E/E_0)^\delta$, 
where $D_0=3\times 10^{28}$ cm$^2$ s$^{-1}$ at $E_0=10$ GeV, and $\chi$ = 1.0 corresponds to the typical value of Galactic diffusion coefficient \citep{2013A&ARv..21...70B}.
For an injected source spectrum given by $Q(E) \propto E^{-\Gamma}$ and $D(E) \propto E^\delta$, the spectrum of escaped protons, $N_p(E)$,
approximately equal $Q(E)$ at low energies where the diffusion radius defined as $r_{\rm diff}=\sqrt{4D(E)T}$ is much smaller than the size of the emission region $R$.
Here $R$ is adopted to be 28.3/6.3~pc with the distance of 0.9/0.2~kpc.
And at high energies, $N_p(E)$ will follow $N_p(E) \propto E^{-\left(\Gamma+\frac{3}{2}\delta\right)}$, 
where the spectral break shown at $E_{\rm p, bre}$ with $R=\sqrt{4D(E_{\rm p, bre})T}$.
With the different parameters of $\eta$, $\chi$ and $\delta$ adopted,
the different spectra of escaped protons in the $\gamma$-ray emission region are produced and the corresponding $\gamma$-ray fluxes are calculated with the $naima$ package \citep{naima}.
And the value of the ambient gas density is assumed to be $n_{\rm gas}$ = 1.0~cm$^{-3}$ considering the absence of the observations of molecular clouds in this region.

\begin{figure}
\centering
\includegraphics[angle=0,scale=0.58]{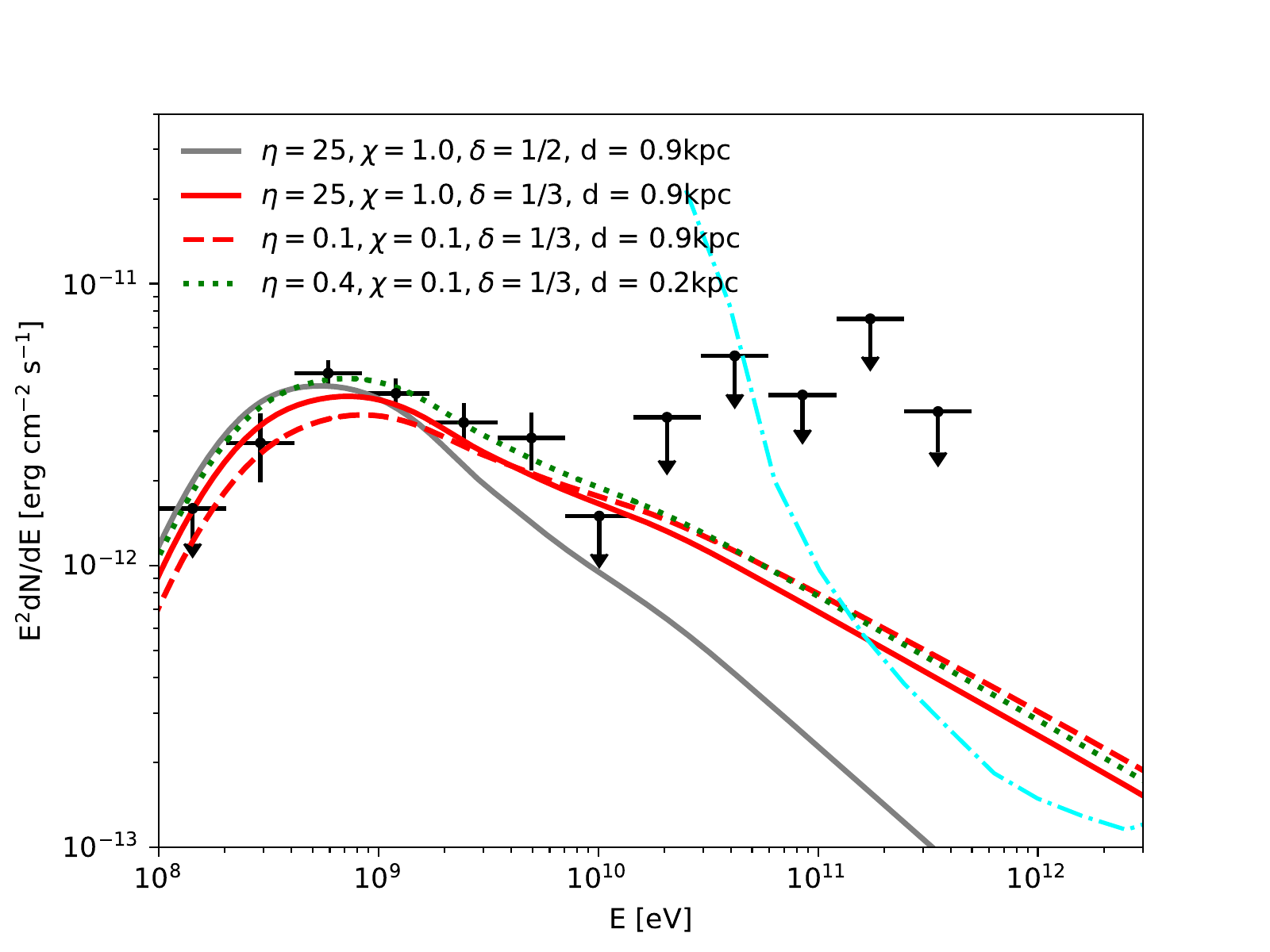}
\caption{Modeling of the $\gamma$-ray spectra with the hadronic escaping model. The different lines indicate the scenarios with the different $\eta$, $\chi$, $\delta$ and distance values as shown in the legend. The cyan dotted–dashed line shows the differential sensitivity of CTA-North \citep[50 hrs;][]{2019scta.book.....C}.}
\label{fig:EscapeModel}
\end{figure}

The resulting hadronic $\gamma$-ray flux with the different parameters are shown in Figure \ref{fig:EscapeModel}. 
Compared with the spectra of $\delta$ = 1/3 with Kolmogorov turbulence for the diffusion coefficient \citep{2006ApJ...642..902P}, the higher value with $\delta$ = 1/2 for Kraichnan turbulence gives a much softer $\gamma$-ray spectrum in the high energy,
which is not very consistent with the observation.
With the typical value of Galactic diffusion coefficient, $\chi$ = 1.0, the total energy of injected protons is fitted to be 2.5 $\times$ 10$^{\rm 51}$ erg, which is not reasonable.
And the total energy of escaped protons above 1~GeV in this region is calculated to be $W_{\rm p}$ = 1.5$\times$ 10$^{49}$ ($n_{\rm gas}$/1.0~${\rm cm^{-3}}$)$^{-1}$~erg.
It should be noted that the much higher total energies could be attributed to the underestimated of the gas density in this region.
In addition, by fixing the total energy of injected protons to be 10$^{\rm 50}$~erg, $\eta$ = 0.1, the $\gamma$-ray spectrum with the distance of 0.9~kpc also can be explained by the hadronic escaping model with a lower diffusion coefficient, and such value is needed to be one order of magnitude lower than the typical Galactic value.
And for the distance of 0.2~kpc, the total energy of injected protons need to be about 4 $\times$ 10$^{\rm 50}$~erg with $\eta$ = 0.4.
And the corresponding total energy of escaped protons in the $\gamma$-ray emission region are estimated to be $W_{\rm p}$ = 1.2$\times$10$^{49}$ ($n_{\rm gas}$/1.0 ${\rm cm^{-3}}$)$^{-1}$~erg and $W_{\rm p}$ = 8.4$\times$10$^{47}$ ($n_{\rm gas}$/1.0 ${\rm cm^{-3}}$)$^{-1}$~erg for the distance of 0.9~kpc and 0.2~kpc, respectively.

\subsection{Pulsar Wind Nebula}
Taking into account the detected pulsar in the $\gamma$-ray emission region, a scenario of the pulsar wind nebula (PWN) driven by PSR J0837-2454 is also considered.
Such extended $\gamma$-ray emissions are also detected in several typical PWNe, such as HESS J1825-137 \citep{2020A&A...640A..76P}, HESS J1640-645 \citep{2018ApJ...867...55X}, etc.
However, these $\gamma$-ray PWNe detected by {\em Fermi}-LAT are driven by the energetic pulsars with spin-down powers between 10$^{\rm 36}$ and 10$^{\rm 39}$ erg s$^{\rm -1}$ \citep{2013ApJ...773...77A}. 
And the spin-down luminosity of PSR J0837-2454 with $\dot{E}$ = $5.5 \times 10^{\rm 34}$ erg s$^{\rm -1}$ seems to be too low to produce such energetic $\gamma$-ray emissions around it.
So we suggest that the PWN scenario for the observed $\gamma$-ray emission for \sg~is not favored.

\subsection{Pulsar Halo}

\begin{figure*}
\centering
\includegraphics[angle=0,scale=0.58]{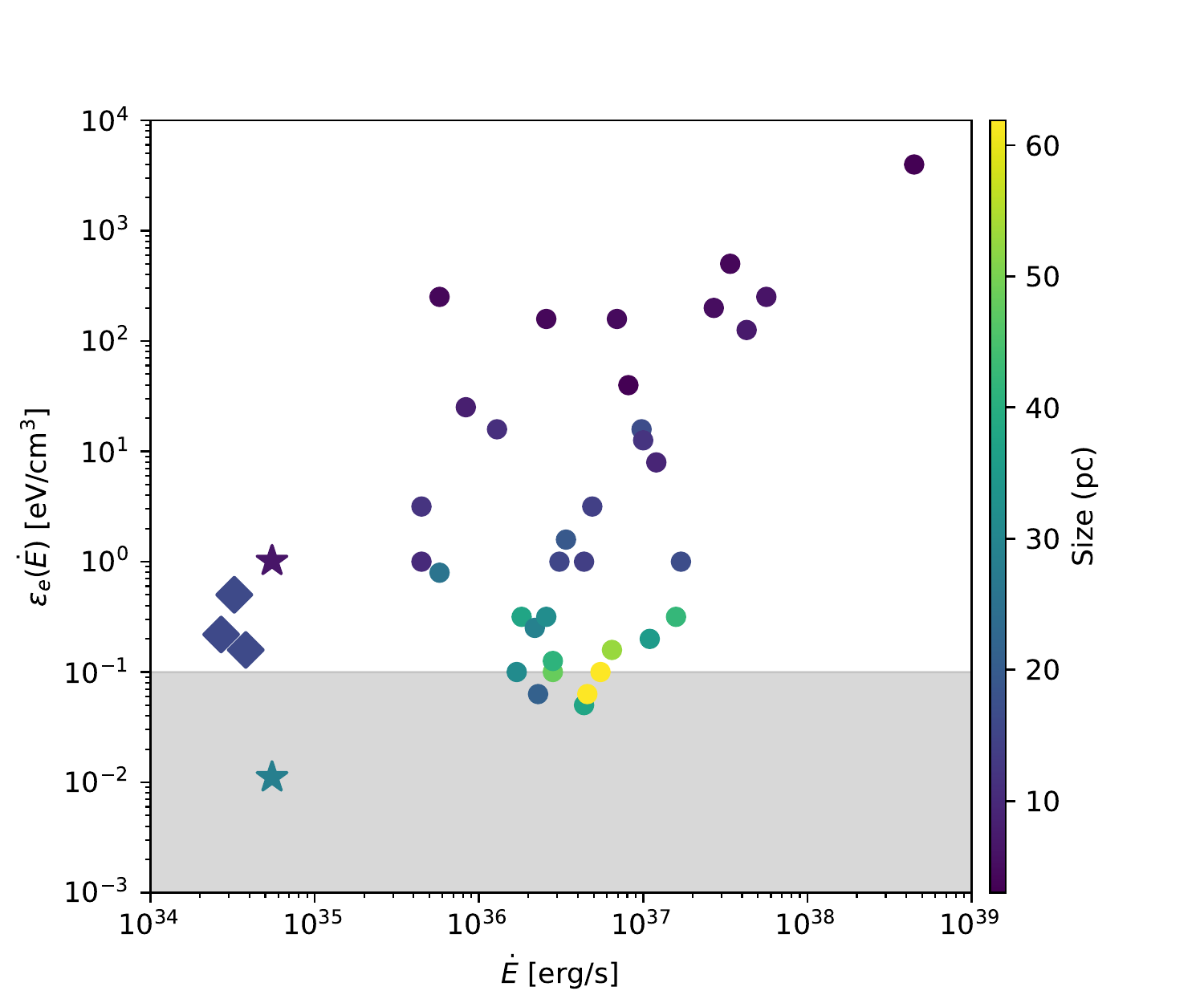}
\includegraphics[angle=0,scale=0.58]{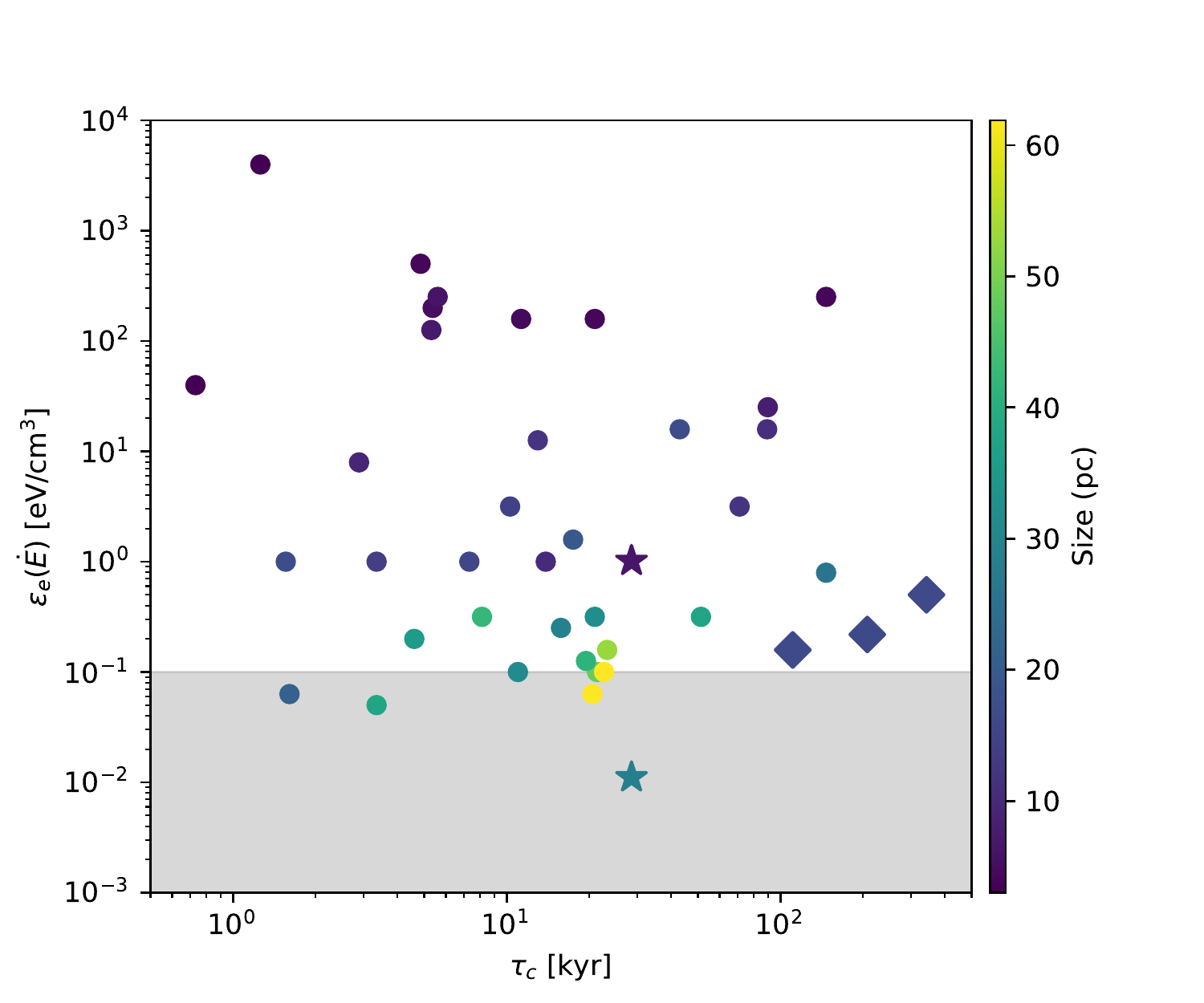}
\caption{Energy density within TeV sources calculated as $\epsilon_{e}$ = $\dot{E}\tau_{\rm c}/V$ as a function of pulsar $\dot{E}$ and $\tau_{\rm c}$ following the Figure 2 in \citet{2020A&A...636A.113G} and the color bar shows the size of the $\gamma$-ray emission of each source. The shaded regions correspond to an energy density lower than that of the ISM of 0.1 eV cm$^{-3}$. The identified pulsar halos associated with Geminga, PSR B0656+14 \citep{2017Sci...358..911A} and PSR J0622+3749 \citep{2021PhRvL.126x1103A} are shown as the diamonds, and the stars represent PSR J0837--2454 with the different distances.}
\label{fig:TeVhalo}
\end{figure*}

Along with the evolution of the PWN into the interstellar medium (ISM), the energetic particles could escape and their transport becomes to be dominated by diffusion. 
And these escaped particles could form a detectable halo around the pulsar, which is defined as a pulsar halo. 
Such halos are first detected as the extended TeV $\gamma$-ray emissions around the nearby low power pulsars Geminga and PSR B0656+14 \citep{2017Sci...358..911A}.
And the searching for the GeV $\gamma$-ray emission for these pulsar halos is ongoing. While only the GeV $\gamma$-ray emission from Geminga halo is detected by {\em Fermi}-LAT \citep{2019PhRvD.100l3015D}, which is still under debate \citep{2019ApJ...878..104X}.
Therefore, we also consider the possible origin as a pulsar halo for \sg~around PSR J0837-2454 that detected here
and discuss the potential TeV observation of it.

For the diffusion process in the pulsar halo, the diffusion size of particles is calculated as $r_{\rm d} = 2\sqrt{D t}$,
and $D$ represents the diffusion coefficient of particles.
Considering the updated distance of 0.9 or 0.2~kpc for PSR J0837-2454 \citep{pbh+21} and the 68\% containment radius of 1$^{\circ}$.8, the physical size of the $\gamma$-ray emission region is calculated to be 28.3 or~6.3 pc.
Adopting the characteristic age of PSR J0837-2454 $\tau_{\rm c}$ = 28.6 kyr, the diffusion coefficient is estimated to be $2 \times 10^{\rm 27}$ cm$^{\rm 2}$/s or 1 $\times 10^{\rm 26}$ cm$^{\rm 2}$/s for d = 0.9 or 0.2~kpc, respectively. 
And such values are much lower than the typical Galactic value of cosmic rays
with $D\simeq3\times10^{28}$~cm$^2$~s$^{-1}$ \citep{2013A&ARv..21...70B}.

Based on the definition of an electron halo in \citet{2020A&A...636A.113G}, namely that of over-density of relativistic electrons around pulsar compared with the ISM, we calculated the energy density $\epsilon_{\rm e}$ in relativistic particles around pulsar with established associated with the $\gamma$-ray emissions with $\epsilon_{\rm e}$ = $E_{\rm inj}/V$,
where $V$ is the volume of the $\gamma$-ray emission region. 
And considering the non-detection of the TeV $\gamma$-ray emission from SG0837, the total injected energy is calculated based on the pulsar properties with $E_{\rm inj}$ = $\dot{E}\tau_{\rm c}$,
where $\dot{E}$ and $\tau_{\rm c}$ are the present spin-down power and characteristic age of the pulsar.
For PSR J0837-2454, the energy density around is estimated to be 0.01 eV cm$^{\rm -3}$ or 1.0 eV cm$^{\rm -3}$
for the distance of $d$ = 0.9 or 0.2~kpc. 
We replot the Figure 2 in \citet{2020A&A...636A.113G} by adding the values of PSR J0837--2454 in our Figure~\ref{fig:TeVhalo},
together with PSR J0622+3749, which is identified to be a pulsar halo by LHAASO \citep{2021PhRvL.126x1103A}.

From Figure~\ref{fig:TeVhalo}, we can see that the energy density around PSR J0837-2454 with $d$ = 0.2~kpc and its current spin-down luminosity are close to the characteristics of TeV halo, like Geminga and PSR B0656+14.
However, the characteristic age of PSR J0837-2454 is at least one order of magnitude lower than that of other halos.
Hence, the extended $\gamma$-ray emission of SG0837 is not much favored for the halo scenario. 
Nonetheless, the potential TeV $\gamma$-ray emission from this source could be expected by the Cherenkov telescopes in the future.



\section{Summary}
\label{sec:sum}

\citet{pbh+21} claimed a discovery and timing for a young pulsar PSR J0837--2454 with
$P$=629.4 ms and $\dot{P}$=3.5$\times10^{-13}$ s s$^{-1}$ by using the radio data from
the Parkes radio telescope. Moreover, an extended low-surface-brightness diffuse emission
around PSR J0837-2454 was also detected by the radio data from the GLEAM, which suggests
it to be a SNR candidate.
Motivated by these, we analyzed the 14 yr $\gamma$-ray data from the \emph{Fermi}-LAT observations
surrounding PSR~J0837--2454. Interestingly, we found a significant extended $\gamma$-ray emission
named as \sg~at a significance level of $\sim12\sigma$ (see Figure~\ref{fig:map}),
which is spatially coincident with the SNR candidate shown as in Figure~\ref{fig:radio}.
And \sg~has a spatial extension with a 68\% containment radius of $\sim1^{\circ}.8$.
The extension significance level is 9.6$\sigma$ with a 2D Gaussian spatial model comparing with a point source model.
And its SED in 0.1--500.0 GeV can be well described by the LP model. 

PSR J0837--2454 is one of the relatively young pulsars compared with other cataloged pulsars.
\citet{pbh+21} had shown 74 pulsars with characteristic age $\leqslant$ 28.6 kyr in their Figure~9
and summarised that 80\% of 74 pulsars are often associated with a SNR and/or PWN.
Considering the above percentage coupled with a diffuse emission in radio
and an extended emission in $\gamma$-rays spatially coincident with the pulsar,
we suggest that \sg~is correlated with the SNR candidate around PSR J0837--2454.
In our spatial analysis, no significant point source was found by subtracting the extended $\gamma$-ray emission from \sg.
And the $\gamma$-ray pulsation of  PSR J0837--2454 was also not found in the timing analysis.
These factors make PSR J0837--2454 to be a radio loud and $\gamma$-ray quiet pulsar.

Several scenarios for the potential origins of the extended $\gamma$-ray emission are discussed,
such as a SNR, PWN, or pulsar halo.
Based on the model fitting results, see the discussion in Section~\ref{sec:dis}, the $\gamma$-ray emission origin of
the \sg~is preferred for the SNR scenario.
And the future potential detection in the TeV band by the Cherenkov Telescope Array in the northern hemisphere \citep[CTA-North;][]{2019scta.book.....C} and the molecular clouds observations could be help to test the different models.

\section*{acknowledgments}
We thank anonymous referee for her/his very helpful suggestions.
This work is supported in part by the National Natural Science Foundation of China
No.~12163006, No.~12233006 and No.~12103040, the Basic Research Program of Yunnan Province
No.~202201AT070137, the joint foundation of Department of Science and
Technology of Yunnan Province and Yunnan University No.~202201BF070001-020, 
and the Natural Science Foundation for Young Scholars of Sichuan Province, China (No.~2022NSFSC1808).


\bibliographystyle{aasjournal}
\bibliography{aas}
\end{document}